\documentclass{ws-procs975x65}
\makeindex
\usepackage{hyperref}
\hypersetup{
    unicode=false,          
    colorlinks=true,       
    linkcolor=black,          
    citecolor=black,        
    filecolor=black,      
    urlcolor=black           
}
\usepackage{enumerate}
\def\d  {\delta}

\def\cL {{\cal L}}

\def\be {\begin{equation}}
\def\ee {\end{equation}}
\def\beq{\begin{equation}}
\def\eeq{\end{equation}}
\def\bea {\begin{eqnarray}}
\def\eea {\end{eqnarray}}
\def\br{\begin{eqnarray}}
\def\er{\end{eqnarray}}
\def\bc {\begin{center}}
\def\ec {\end{center}}
\def\bfg {\begin{figure}}
\def\efg {\end{figure}}
\def\bi {\begin{itemize}}
\def\ei {\end{itemize}}
\def\benu{\begin{enumerate}}
\def\eenu{\end{enumerate}}
\newcommand{\bdm}{\begin{displaymath}}
\newcommand{\edm}{\end{displaymath}}
\def\laq{\hbox{~}\raise 0.4ex\hbox{$<$}\kern -0.8em\lower 0.62ex\hbox{$\sim$}\hbox{~}}
\def\gaq{\hbox{~}\raise 0.4ex\hbox{$>$}\kern -0.7em\lower 0.62ex\hbox{$\sim$}\hbox{~}}
\newsavebox{\blambox}\savebox{\blambox}[0.6em]{$\lambda\!\!\!$\raisebox{0.5em}
{$\neg$}}
\newsavebox{\bFox}\savebox{\bFox}[0.6em]{$F\!\!\!\!$\raisebox{0.5em}
{$\neg$}}
\newsavebox{\bxibox}\savebox{\bxibox}[0.6em]{$\xi\!\!\!$\raisebox{.5em}
{$\neg$}}
\newcommand{\pbg}{\phi_{_0}}
\newcommand{\dpbg}{\dot\phi_{_0}}

\newcommand{\PXbg}{P_{_X}^{^{(0)}}}

\newcommand{\PXXbg}{P_{_{XX}}^{^{(0)}}}

\newcommand{\PXXXbg}{P_{_{XXX}}^{^{(0)}}}

\newcommand{\dophi}{\delta\phi}
\newcommand{\ddophi}{\delta\dot{\phi}}

\newcommand{\dii}{\delta^{^{(2)}}\!}
\begin{document}
\title{Ambiguities in second-order cosmological perturbations for
non-canonical scalar fields}
\author{Corrado~Appignani$^{(1)}$,
Roberto~Casadio$^{(1)}$,
S.~Shankaranarayanan$^{(2,3)}$\footnote{speaker; E-mail:~shanki@iisertvm.ac.in}}
\address{${~}^{(1)}$ Dipartimento di Fisica, Universit`a di Bologna
and
I.N.F.N., Sezione di Bologna, via Irnerio 46, 40126 Bologna, Italy
\\
${~}^{(2)}$ Institute of Cosmology and Gravitation, University of 
Portsmouth, Portsmouth, UK
\\
${~}^{(3)}$ School of Physics, Indian Institute of Science Education 
and Research-Trivandrum,
\\
CET campus, Thiruvananthapuram 695 016, India}
\begin{abstract}
Over the last few years, it was realised that non-canonical
scalar fields can lead to the accelerated expansion in the early
universe.
The primordial spectrum in these scenarios not only shows
near scale-invariance consistent with CMB observations,
but also large primordial non-Gaussianity.
Second-order perturbation theory is the primary theoretical tool
to investigate such non-Gaussianity.
However, it is still uncertain which quantities are gauge-invariant
at second-order and their physical understanding therefore
remains unclear.
As an attempt to understand second order quantities, we consider a
general non-canonical scalar field, minimally coupled to gravity,
on the unperturbed FRW background where metric fluctuations
are neglected a priori.
In this simplified set-up, we show that there arise ambiguities in the
expressions of physically relevant quantities, such as the effective
speeds of the perturbations.
Further, the stress tensor and energy density display a potential
instability which is not present at linear order.
\end{abstract}
\bodymatter
\section{Introduction and motivation}
Predictions of inflation seem to be in excellent agreement with the
CMB data~\cite{2009-Komatsu.etal-ApJS}.
However, it is still unclear what is the nature of the field which drives
inflation.
Historically, a canonical scalar field has been the preferred candidate
for inflaton, but in recent years, also a non-canonical scalar field,
dubbed as $k$-inflaton, was considered as serious alternative
mechanisms to drive inflation~\cite{1999-Armendariz-Picon.etal-PLB}.
\par
Both scenarios lead to nearly scale invariant power-spectra with
negligible running and hence can not be distinguished (or ruled out)
from the current CMB data~\cite{2009-Komatsu.etal-ApJS}.
The future missions, including PLANCK~\cite{2005-Planck-ScientificProgrammeof},
hold promise in ruling our either of these two scenarios by looking at
the non-Gaussianity of the primordial spectra, but
quantifying non-Gaussianity requires one to go beyond linear
order~\cite{1997-Bruni.etal-CQG,2003-Maldacena-JHEP,2005-Seery.Lidsey-JCAP}.
\par
There are four different approaches in the literature to study
cosmological perturbations:
{\tt 1)}~solving Einstein's equations
order-by-order~\cite{1963-Lifshitz.Khalatnikov-AdP};
{\tt 2)}~the covariant approach based on a general frame vector
$u^{\alpha}$~\cite{1966-Hawking-ApJ};
{\tt 3)}~the Arnowitt-Deser-Misner (ADM) approach based on the normal 
frame vector $n^{\alpha}$~\cite{1980-Bardeen-PRD};
{\tt 4)}~the reduced action approach~\cite{1980-Lukash-ZhETF}.
In the case of linear perturbations, it was shown that all of
these four approaches lead to identical equations of motion.
However, to our knowledge, a complete analysis has not been done in the
literature for higher-order perturbations.
\par
In this talk, to illustrate the problems that may occur at
higher-order, and not to get bogged-down with the gauge issues, we
consider a simple situation:
we freeze all metric perturbations
and focus on the perturbations of a minimally-coupled,
generalised scalar field $\phi$, whose Lagrangian density is given
by~\cite{1999-Armendariz-Picon.etal-PLB}
{\small 
\beq
\label{eq:genscal}
\cL = P(X, \phi) \ , \qquad \mbox{where} \qquad 2\,X = \nabla^{\alpha}
\phi\,\nabla_{\alpha} \phi \ .  
\eeq } 
More precisely, we will only consider linear perturbations of the
scalar field,
{\small \beq
\label{eq:def-Perphi}
\phi(t, {\bf x})=\pbg(t) + \d^{(1)}\phi(t, {\bf x})
\ ,
\eeq
}
about the 4-dimensional FRW~background, while expanding all the
dependent quantities, like $X$ and stress tensor, up to second order,
and highlight the main differences in these approaches.
For this purpose, it is convenient to compare the ratio
{\small	
\beq
\label{eq:cs-def}
c_{\rm s}^2 = \frac{\mbox{coefficients of~} (\dophi_{,i}/a)^2}
{\mbox{coefficients of~}
\ddophi^2}
\ ,
\eeq
}
in the components of the stress~tensor and related quantities.
Since $c_{\rm s}^2$ is dimensionally the square of a speed,
we will refer to this ratio as the ``speed of propagation''.
\section{Key results}
Below, we will provide the key results and, for details, we refer the
reader to Ref.~\cite{2009-Appignani.etal-Arx}.
\\[5pt]
\noindent
{\tt Perturbed tensor and ADM approach:}
For an arbitrary scalar field Lagrangian, $\dii T_{00}$ may represent
an unstable perturbation; only under very special conditions, most notably
for the canonical scalar field, the {\it effective speed of propagation\/} of
$\dii T_{00}$ (and $\dii T_{ii}$) are the same as that of the standard
definition~\cite{1999-Garriga.Mukhanov-PLB}
{\small	
\beq
c^2_{\dophi} = \frac{\PXbg}{\PXbg  +  \PXXbg\,\dpbg^2}
\eeq
}
Using~\eqref{eq:cs-def}, we can define speeds related with the
propagation of density perturbations in the background frame from
$\dii T_{00}$ and $\dii T_{ii}$, respectively, 
{\small 
\beq
\label{eq:csT}
  c_{0}^2= \strut\displaystyle\frac{\PXbg - \PXXbg\, \dpbg^2} {\PXbg
    +4\,\PXXbg\,\dpbg^2+ \PXXXbg\, \dpbg^4} \, , \quad c_{\parallel}^2
  = c^2_{\delta \phi} \, .
\eeq
}
Note that these velocities become imaginary indicating that the
perturbations may be unstable.
The nature of this instability is easily understood in analogy
with classical mechanics:
when $c_{\dophi}^2$ is negative, the system resembles an inverted
harmonic oscillator and, no matter how small $\delta \phi$,
it will rapidly run away from the background solution $\pbg$ and from the
perturbative regime.
\\[5pt]
\noindent
{\tt Covariant approach:}
In the fluid frame, the energy density exhibits the same kind of instability
as the perturbed stress-tensor $\dii T_{00}$, also for the canonical scalar field.
In particular, the perturbations turn out to be unstable because of the negative
sign of the spatial momentum contribution.
Only under special conditions, the {\it effective speed of propagation\/}
of the energy density and pressure perturbations are equal.
Using~\eqref{eq:cs-def}, the speed of propagation for density and pressure
perturbations in the fluid frame are 
{\small
\beq 
c_{\rho}^2 = - \frac{\PXbg + \PXXbg\,\dpbg^2} {\PXbg
  +4\,\PXXbg\,\dpbg^2 + \PXXXbg\, \dpbg^4} \, , \quad
c_{p}^2=c_{\parallel}^2=c_{\delta\phi}^2
\ , 
\ee 
}
which differ from Eq.~\eqref{eq:csT}.
\\[5pt]
\noindent
{\tt Symmetry reduced approach:}
The basic idea here is to perturb the action about the FRW~background,
up to second (or higher) order, and reduce it so that the perturbations are
described in terms of a single gauge-invariant variable.
The second order canonical Hamiltonian is identical to the stress-tensor
for the canonical scalar field, but differs for general non-canonical fields.
This implies that $\dii T_{00}$ and the canonical Hamiltonian
$\dii\mathcal{H}$ may become unstable under different conditions.
\par
\section{Discussion} 
The first question is why the canonical Hamiltonian, perturbed
stress~tensor and super-Hamiltonian coincide for a canonical scalar
field, but not for general scalar field Lagrangians.
To go about answering this question, it is necessary to look at the four
approaches we have employed from a different perspective.
In the first two approaches -- perturbed stress~tensor and covariant
approach -- we perturb the general expression for the scalar field
stress~tensor and obtain its second order contribution $\dii T_{00}$.
In the last two approaches -- ADM~formulation and symmetry-reduced
action -- we expand the action to second order in the perturbation and
obtain the super-(canonical) Hamiltonian of the corresponding
perturbed action.
While the super-Hamiltonian $\dii H$ is identical to $\dii T_{00}$,
the canonical Hamiltonian $\dii\mathcal{H}$ differs.
\par
Under what condition the super-Hamiltonian $\dii H$, the canonical
Hamiltonian $\dii\mathcal{H}$ for non-canonical scalar fields are
identical?
They are all identical provided the time-variation of
background quantities (like $\PXbg$ and $\PXXbg$) can be neglected.
(For the canonical scalar field, these functions are indeed constant
and the perturbed quantities therefore coincide.)
Although such an approximation may be valid for specific non-canonical
fields, they fail for some of the known fields like Tachyon,
DBI~\cite{2009-Appignani.etal-Arx}.
\par
R.~C.~is supported by the INFN grant BO11 and S.~S.~was supported by
the Marie Curie Incoming International Grant IIF-2006-039205.

\begin{thebibliography}{10}
%
\bibitem{2009-Komatsu.etal-ApJS}
E.~Komatsu et al,
{\em Astrophys. J. Sup.} {\bf 180}, 330 (2009).
%
\bibitem{1999-Armendariz-Picon.etal-PLB}
C.~Armendariz-Picon, T.~Damour and V.~F. Mukhanov, {\em Phys. Lett.} {\bf
B458}, 209 (1999).
%
\bibitem{2005-Planck-ScientificProgrammeof}
Planck, {\em The Scientific Programme of Planck} (ESA Publications, 2005).
%
\bibitem{1997-Bruni.etal-CQG}
M.~Bruni et al,
{\em Class. Quant. Grav.} {\bf 14}, 2585 (1997).
%
\bibitem{2003-Maldacena-JHEP}
J.~M.~Maldacena, {\em JHEP} {\bf 05}, p. 013 (2003).
%
\bibitem{2005-Seery.Lidsey-JCAP}
D.~Seery and J.~E.~Lidsey, {\em JCAP} {\bf 0506}, p. 003 (2005).
%
\bibitem{1963-Lifshitz.Khalatnikov-AdP}
E.~M.~Lifshitz and I.~M.~Khalatnikov, {\em Adv. Phys.} {\bf 12}, 185 (1963).
%
\bibitem{1966-Hawking-ApJ}
S.~W.~Hawking, {\em Astrophys. J.} {\bf 145}, 544 (1966).
%
\bibitem{1980-Bardeen-PRD}
J.~M.~Bardeen, {\em Phys. Rev.} {\bf D22}, 1882 (1980).
%
\bibitem{1980-Lukash-ZhETF}
V.~N.~Lukash, {\em ZhETF} {\bf 31}, 631 (1980).
%
\bibitem{2009-Appignani.etal-Arx}
C.~Appignani, R.~Casadio and S.~Shankaranarayanan, arxiv: 0905.4184 (2009).
%
\bibitem{1999-Garriga.Mukhanov-PLB}
J.~Garriga and V.~F.~Mukhanov, {\em Phys. Lett.} {\bf B458}, 219 (1999).
\end{thebibliography}
\end{document}